\documentclass{article}
\usepackage{epsfig}
\usepackage{amsmath}
\newtheorem{theorem}{Theorem}[section]
\newtheorem{lemma}{Lemma}[section]
\newtheorem{corollary}{Corollary}[section]

\newtheorem{definition}{Definition}[section]
\newcommand{\blackslug}{\penalty 1000\hbox{
    \vrule height 8pt width .4pt\hskip -.4pt
    \vbox{\hrule width 8pt height .4pt\vskip -.4pt
          \vskip 8pt
      \vskip -.4pt\hrule width 8pt height .4pt}
    \hskip -3.9pt
    \vrule height 8pt width .4pt}}

\newenvironment{proof}{\vspace{1mm} \noindent {\sc Proof.}$\;$\rm}{\qed}
\newcommand{\qed}{\hspace*{\fill}\blackslug}
\def\boxit#1{\vbox{\hrule\hbox{\vrule\kern4pt
 \vbox{\kern1pt#1\kern1pt}
\kern2pt\vrule}\hrule}}
\begin{document}

\title{\bf Time and Space Efficient Algorithms for RNA Folding with the Four-Russians Technique}

\author{Yinglei Song\\
School of Electronics and Information Science\\
Jiangsu University of Science and Technology\\
Zhenjiang, Jiangsu 212003, China\\
Email: syinglei2013@163.com\\
}
\date{}
\maketitle

\begin{abstract}
\noindent In this paper, we develop new algorithms for the basic RNA folding problem. Given an RNA sequence that contains $n$ nucleotides, the goal of the problem is to compute a pseudoknot-free secondary structure that maximizes the number of base pairs in the sequence. We show that there exists a dynamic programming algorithm that can solve the problem in time $O(\frac{n^{3}}{\log_{2}{n}})$ while using only $O(\frac{n^{2}}{\log_{2}{n}})$ memory space. In addition, we show that the time complexity of this algorithm can be further improved to $O(\frac{n^{3}}{\log_{2}^{2}{n}})$ at the expense of a slightly increased space complexity. To the best of our knowledge, this is the first algorithm that can solve the problem with traditional dynamic programming techniques in time $O(\frac{n^{3}}{\log_{2}^{2}{n}})$. In addition, our results improve the best known upper bound of the space complexity for efficiently solving both this problem and the context-free language recognition problem.
\end{abstract}

\section{Introduction}

An important problem in structural bioinformatics is to determine the secondary structure of an RNA molecule from its primary sequence of nucleotides \cite{nussinov, nussinov1, waterman, zuker}. Although the problem was originally proposed more than thirty years ago, it is still under intensive research and remains to be an important topic in structural bioinformatics. The problem has important applications in many areas of molecular biology including RNA structure prediction \cite{nussinov, nussinov1}, RNA design \cite{siederdissen} and the study of complete genomes of RNA viruses \cite{hofacker}. An algorithm that can efficiently and accurately determine the secondary structure of an RNA sequence is thus highly desirable for researchers in these areas.

Given an RNA sequence that contains $n$ nucleotides,
the goal of the basic RNA folding problem is to find a
secondary structure that does not contain crossing base pairs
and maximizes the number of base pairs in the sequence.
The problem can be solved with a well established
dynamic programming algorithm in $O(n^{3})$ time and
$O(n^{2})$ memory space \cite{durbin}.

Since this algorithm is available, researchers have made a large amount of efforts to further reduce the time and space
complexities of the problem. In \cite{backofen, wexler}, a few heuristics are developed to reduce the
computation time of the dynamic programming algorithm in practice. In \cite{akutsu}, an algorithm that needs
$O(\frac{n^{3}\log_{2}\log_{2}n}{\log_{2}^{\frac{1}{2}}n})$ time is developed to solve the problem. However, the algorithm is complex and difficult to implement. In \cite{backofen}, an additional parameter $Z$ is introduced to describe an instance of the problem. $Z$ is an integer between $n$ and $n^{2}$ and the time and space complexities of the problem are shown to be $O(nZ)$ and $O(Z)$ respectively. However, the worst-case time and space complexities of the algorithm remain to be $O(n^3)$ and
$O(n^2)$. Although it is observed in the same paper that the time complexity of the algorithm can be
improved to $O(\frac{n^{3}\log_{2}^{3}{\log_{2}{n}}}{\log_{2}^{2}{n}})$ by combining it with a new algorithm for computing all pairs of shortest paths in a graph \cite{chan}, this combined algorithm is difficult to implement and its practical value is thus limited.

Recently, significant progress has been made toward improving the worst-case time complexities of the basic RNA folding problem and a few other related problems. In \cite{pinhas}, a faster algorithm for matrix-vector min-plus multiplication is developed based on a certain discreteness assumption. It is shown that the worst-case time complexity of the basic RNA folding problem can be improved to $O(\frac{n^{3}}{\log_{2}^{2}{n}})$ by combining this algorithm with the Valiant's technique \cite{valiant}. In \cite{zakov}, a few problem templates are developed and generic algorithms using Valiant's technique are developed for problems that can be fit into these templates. Improved time complexity results for a few related problems are obtained based on this observation.

On the other hand, new techniques have been developed to improve the computation time needed to solve the problem using only traditional dynamic programming techniques. In \cite{frid, venk}, it is shown that the problem can be solved in time $O(\frac{n^3}{\log_{2}n})$ with the Four-Russians technique. Four-Russians technique is an elegant approach that has been successfully used to reduce the worst-case time complexity of dynamic programming algorithms. For example, an early application of the technique \cite{masek} successfully reduces the worst-case time complexity of the longest common sequence problem from $O(n^2)$ to
$O(\frac{n^{2}}{\log_{2}^{2}{n}})$. Although it appears to be difficult to further reduce the time and space complexities of the dynamic programming approach without using Valiant's technique. The successful application of the Four-Russians technique to the problem provides a new perspective on solving the problem with less computational resources. A question therefore arises naturally on whether the Four-Russians technique can be utilized to further reduce the amount of computation time and space needed to solve the basic RNA folding problem with only traditional dynamic programming techniques.

Compared with the Valiant's technique, traditional dynamic programming techniques are easier to implement in practice and can be easily extended to solve the RNA folding problem when the secondary structure may contain pseudoknots. It is therefore highly desirable to develop algorithms that are based on traditional dynamic programming techniques and also efficient in both time and space.

In this paper, we develop new techniques to further utilize the Four-Russians technique to improve the computation time and space needed to solve the problem with traditional dynamic programming techniques. We start by showing that the problem can be solved with a dynamic programming algorithm in time $O(\frac{n^{3}}{\log_{2}{n}})$ while using only $O(\frac{n^{2}}{\log_{2}{n}})$ memory space. We then show that the time complexity of the algorithm can be further improved to $O(\frac{n^{3}}{\log_{2}^{2}{n}})$ with a slightly increased space complexity. To the best of our knowledge, this algorithm is the first algorithm that can solve the problem in time $O(\frac{n^{3}}{\log_{2}^{2}{n}})$ without using Valiant's technique. In addition, our results improve the known worst-case space complexity of the basic RNA folding problem, which is $O(n^2)$. One important advantage of our algorithms is that they are not difficult to implement. Another important observation is that our techniques can also be used to improve the worst-case space complexity needed to efficiently solve the context-free language recognition problem, which has been unchanged for more than thirty years.

\section{Preliminaries and Notation}

In general, an RNA molecule consists of a sequence of nucleotides. There are four types of nucleotides and we use letters A, C, G, U to denote them. In addition, we use $\Sigma=\{A,C,G,U\}$ to denote the set of all possible nucleotides a sequence may contain, an RNA sequence is considered to be a string formed by letters from $\Sigma$. A {\it base pair} is a pair of nucleotides that interact and form strong hydrogen bonds in the structure of an RNA sequence.

Molecular biology has revealed that energetically stable base pairs are usually formed between A and U or G and C. Two nucleotides are {\it complimentary} if they can form an energetically stable base pair. Without loss of generality, we only consider these two possible types of base pairs in this paper. The {\it secondary structure} of an RNA sequence is comprised of a set of base pairs formed by nucleotides in the sequence. Two base pairs in a sequence {\it cross} each other if they are neither nested nor in parallel in the sequence. Since the secondary structures of most RNA sequences do not contain crossing base pairs, the goal of the basic RNA folding problem is to compute a secondary structure that contains the maximum number of non-crossing base pairs.

Given an RNA sequence,
two nucleotides that form a base pair are usually separated by at least a certain number of nucleotides along the sequence.
To simplify the presentation of our algorithm, we assume that any pair of complimentary nucleotides in a sequence can form a base pair. It is not difficult to see later that our algorithm can be slightly modified to guarantee the distance
restriction for base pairs while maintaining its time and space complexities.

The well known $O(n^3)$ time dynamic programming algorithm for this problem is similar to the CYK algorithm for context free language recognition. Specifically, given a sequence $S[1\cdots n]$, we maintain a $n \times n$ table $M$ such that $M[i, j]$ stores the maximum number of non-crossing base pairs that can be formed in nucleotides in subsequence $S[i\cdots j]$, where $1 \leq i \leq j \leq n$. Given two nucleotides $a$ and $b$, $C[a, b]=1$ if $a$ and $b$ are complimentary, otherwise $C[a, b]=0$. The recursion relation for computing $M[i, j]$ is as follows.
\begin{equation}
M[i, j]=\max{\{M[i-1, j], M[i, j-1], M[i-1,j-1]+C[S[i], S[j]], N[i, j]\}}
\end{equation}
where $S[i]$ and $S[j]$ are the nucleotides in positions $i$ and $j$ in $S$ and $N[i, j]$ can be computed as follows.
\begin{equation}
N[i, j]=\max_{i \leq p <j}{\{M[i, p]+M[p+1, j]\}}
\end{equation}

It is not difficult to see that the computation of $N[i, j]$ may need $O(n)$ time and the worst-case time complexity of the algorithm is thus $O(n^3)$. Since the algorithm needs to maintain an $n \times n$ table $M$ to store the intermediate computational results, its space complexity is $O(n^2)$. It is clear that the computation of $N[i, j]$ and the storage of each $M[i, j]$ are the obstacles for us to further reduce its worst-case time and space complexities respectively, our major goal in the rest of the paper is therefore to develop methods that can compute the value of $N[i, j]$ more efficiently while using less amount of memory space.

We need to point out that the above analysis of the dynamic programming algorithm is based on the assumption that any positive integer that is at most $n$ can be stored in one word of the computer system. Since otherwise, a single arithmetic step may need up to $O(\log_{2}{n})$ time and the worst-case time complexity of the algorithm increases from the original $O(n^3)$ to $O(n^{3}\log_{2}n)$. The same assumption is in fact also implied in the time complexity results obtained in most of the related work \cite{akutsu, frid, zakov}. In the rest of the paper,  we refer to this assumption as the {\it word width assumption} and the analysis of the time complexity of our algorithm is also based on this assumption.

\section{The Memory Efficient Algorithm}
\label{sec0}
In this section, we show how the Four-Russians technique can be utilized to develop an algorithm that can solve the problem in $O(\frac{n^{3}}{\log_{2}{n}})$ time and $O(\frac{n^{2}}{\log_{2}{n}})$ memory space. In section \ref{sec1}, we show how the time complexity of the algorithm can be further improved to $O(\frac{n^{3}}{\log_{2}^{2}n})$. Although the basic idea of our algorithm is similar to that of the algorithm developed in \cite{frid}, additional techniques are used to reduce the space complexity of the problem from $O(n^{2})$ to $O(\frac{n^{2}}{\log_{2}{n}})$.

The basic idea of the Four-Russians technique is to divide the dynamic programming table into parts and precompute the possible contents in each part. The precomputation step is usually performed before the dynamic programming procedure starts and the results of the precomputation are stored in {\it query tables}. During the dynamic programming procedure, the content of each part can be efficiently obtained by querying the query tables with the information that has been previously obtained. In \cite{frid}, it has been shown that precomputation and the dynamic programming procedure can be effectively interleaved and the time complexity of the dynamic programming procedure can be reduced from $O(n^3)$ to $O(\frac{n^3}{\log_{2}{n}})$.

Given an RNA sequence $S[1\cdots n]$ that contains $n$ nucleotides and integers $i$ and $j$ such that
$1\leq i \leq j \leq n$, we use $M[i, j]$ to denote the maximum number of non-crossing base pairs that can be formed in subsequence $S[i\cdots j]$. The following lemma establishes the foundation of our algorithm. Although the fact stated in this lemma has been used in \cite{frid} as an obvious fact to develop the $O(\frac{n^3}{\log_{2}{n}})$ time algorithm, we provide a formal proof for completeness.
\begin{lemma}
\rm
\label{lm1}
Let $1 \leq i \leq j \leq n$, where $n$ is the length of the sequence $S$, the following inequalities hold.
\begin{enumerate}
\item{$M[i, j] \leq M[i, j+1] \leq M[i, j]+1$}
\item{$M[i+1, j] \leq M[i, j] \leq M[i+1, j]+1$}
\end{enumerate}

\begin{proof}
We let $l=M[i, j]$ and we use $S_{i, j}$ to denote the set of $l$ non-crossing base pairs that are formed in subsequence $S[i\cdots j]$. Similarly, we let
$m=M[i, j+1]$ and $S_{i, j+1}$ to denote the set of $m$ non-crossing base pairs that are formed among nucleotides in subsequence $S[i \cdots j]$. If $S_{i, j+1}$ contains a base pair $p$ that forms between a nucleotide in $S[i\cdots j]$ and
$S[j+1]$, $S_{i, j+1}-\{p\}$ is a set of non-crossing base pairs formed among nucleotides in $S[i\cdots j]$. Since $M[i, j]$ is the largest number of non-crossing base pairs that can be formed in $S[i\cdots j]$, we immediately obtain
\begin{equation}
m-1 \leq l
\end{equation}
On the other hand, $S_{i,j}$ is also a set of base pairs formed in $S[i\cdots j+1]$, we immediately obtain the following inequality from the definition of $M[i, j+1]$.
\begin{equation}
l \leq m
\end{equation}
We therefore can conclude that
\begin{equation}
M[i, j] \leq M[i, j+1] \leq M[i, j]+1
\end{equation}
It is straightforward to see that inequality
\begin{equation}
M[i+1, j] \leq M[i, j] \leq M[i+1, j]+1
\end{equation}
holds for the same reason. The lemma thus follows.
\end{proof}
\end{lemma}

We partition the interval $[1\cdots n]$ into $k=\lfloor \frac{n}{w} \rfloor +1$ disjoint intervals and each interval contains up to $w$ consecutive integers, where $w$ is an integer such that $0<w \leq \log_{2}{n}$. The value of $w$ is determined later in the paper. At least $k-1$ of these intervals are of equal length and each of them contains $w$ consecutive integers and the remaining one contains at most $w$ consecutive integers. Specifically, we use $I_s$ to denote the interval $[sw+1, (s+1)w]$ for $0 \leq s < k-1$ and $I_{k-1}$ to denote the last interval $[(k-1)w+1, n]$. Each of these intervals is an {\it aggregate interval}.

Given integers $i$ and $s$ such that $1 \leq i < sw+1  \leq n$, $M[i, sw+1], M[i, sw+2], \cdots, M[i, (s+1)w]$ form an {\it $(i, s)$ left vector}. From Lemma \ref{lm1}, an $(i, s)$ left vector can be completely determined by $M[i, sw]$ and an $s$ dimensional binary vector $l(i, s)$ that only contains $0$ or $1$ in its components. $l(i,s)$ is an {\it $(i, s)$ left difference vector}.

Similarly, given integers $j$ and $s$ such that $1 \leq (s+1)w <  j \leq n$, $M[sw+1, j], M[sw+2, j], \cdots, M[(s+1)w, j]$ form a {\it $(j, s)$ right vector}. From Lemma \ref{lm1}, a $(j, s)$ right vector can be completely determined by $M[sw+1, j]$ and an $s$ dimensional binary vector $r(j, s)$ that only contains $0$ or $1$ in its components. $r(j, s)$ is a {\it $(j, s)$ right difference vector}. A binary vector is a {\it difference vector} if there exists an integer pair $(i, s)$ such that it is an $(i, s)$ left difference vector or an $(i, s)$ right difference vector.

A difference vector can be considered to be the binary encoding of an integer $u$ such that $0 \leq u \leq 2^{w+1}-1$. Since $w < \log_{2}{n}$, $u$ can be stored in one word due to the word width assumption. Since there are in total $O(nk)$ difference vectors and each is associated with a pair of integers. We can store the integers encoded by them in two query tables that can be efficiently queried during the dynamic programming procedure.

\begin{definition}
\rm
Let $S$ be an RNA sequence of length $n$. A {\it left table} $L_{t}$ is an $n \times (k-1)$ table such that $L_{t}[i,s]$ stores the value encoded by the $(i, s)$ left difference vector associated with $S$.
\end{definition}

\begin{definition}
\rm
Let $S$ be an RNA sequence $S$ of length $n$. A {\it right table} $R_{t}$ is an $n \times (k-1)$ table such that $R_{t}[j, s]$ stores the value encoded by the $(j, s)$ right difference vector associated with $S$.
\end{definition}

$L_t$ and $R_t$ are the two tables that store the intermediate results that have been obtained during the dynamic programming procedure. However, they are not the tables that need to be computed during the precomputation stage of the Four-Russians technique. To clarify the goal of the precomputation stage, we need the following definitions.

\begin{definition}
\rm
Let $S$ be an RNA sequence of length $n$, $I_0, I_1, \cdots, I_{k-1}$ are the aggregate intervals of length $w$ in $S$. A {\it central table} $C_{t}$ is an $2^{w+1} \times 2^{w+1}$ table such that $C_{t}$ can be accessed based on two integers $u$ and $v$ that satisfie $0 \leq u < 2^{w+1}$, $0 \leq v < 2^{w+1}$. $C_t[u, v]$ stores two positive integers whose values are at most $w$.
\end{definition}

\begin{definition}
\rm
Let $S$ be an RNA sequence of length $n$, $i$, $j$ and $s$ are
integers that satisfy $0 \leq s < k-1$, $1 \leq i \leq sw$ and $(s+1)w < j \leq n$. The {\it $(i, j)$ representative} of $I_s$ is an integer $r$ such that the following two conditions hold.
\begin{enumerate}
\item{$1 \leq r \leq w$;}
\item{for all $p$'s such that $sw+1 \leq p \leq (s+1)w$, $M[i, p]+M[p+1, j]$ is maximized when $p=ws+r$.}
\end{enumerate}
\end{definition}

\begin{definition}
\rm
Let $S$ be an RNA sequence of length $n$, $i$, $j$ and $s$ are
integers that satisfy $0 \leq s < k-1$, $1 \leq i \leq sw$ and $(s+1)w < j \leq n$. $r$ is the $(i, j)$ representative of $I_s$. The {\it $(i, j)$ deviation} of $I_s$ is the value of $M[i,r]+M[r+1,j]-M[i,sw+1]-M[sw+2,j]$.
\end{definition}

Lemma \ref{lm2} shows that a central table $C_t$ can be precomputed such that both the $(i, j)$ representative and deviation of an aggregate interval $I_s$ can be obtained in $O(1)$ time by querying $C_t$ with the two integers encoded by the $(i, s)$ left difference vector and $(j, s)$  right difference vector respectively.

\begin{lemma}
\rm
\label{lm2}
Let $S$ be an RNA sequence of length $n$, $i$, $j$ and $s$ are integers that satisfy $1 \leq i \leq sw$ and $(s+1)w \leq j \leq n$. Both the $(i, j)$ representative and $(i, j)$ deviation of $I_s$ can be determined by the $(i, s)$ left difference vector and $(j, s)$ right difference vector.

\begin{proof}
We use $U$ to denote the $(i, s)$ left difference vector and
$V$ to denote the $(j, s)$ right difference vector. For $1 \leq q \leq w$, $u(q)$ and $v(q)$ are the $q$th components of $U$ and $V$ respectively. $U(q)$ is the value of $M[i,sw+q]-M[i,sw+q-1]$ and $V(q)$ is the value of $M[sw+q, j]-M[sw+q+1,j]$.
From Lemma \ref{lm1}, the value of $U(q)$ or $V(q)$ is either $0$ or $1$.

Let $T=M[i,sw]+M[sw+1,j]$, it is not difficult to see that
the following equality holds for $1 \leq t \leq w$.
\begin{equation}
M[i, sw+t]+M[sw+t+1, j]=T+\sum_{c=1}^{t}{(U(c)-V(c))}
\end{equation}

Both the $(i, j)$ representative and deviation of $I_s$ can thus be found based on the components of $U$ and $V$. The lemma thus follows.
\end{proof}
\end{lemma}

From Lemma \ref{lm2}, both the $(i, j)$ representative and deviation of $I_s$ only depend on the left and right difference vectors associated with $(i, s)$ and $(j,s)$. The two vectors are both $w$-dimensional binary vectors. In the precomputation stage, our algorithm thus enumerates all possible combinations of two $w$-dimensional binary vectors and computes the representative and deviation associated with each possible combination and store their values in the central table.

Specifically, the algorithm constructs a central table $C_t$ and enumerates all possible combinations of two $w$-dimensional binary vectors $U$ and $V$. For each such combination, we use $e_u$ and $e_v$ to denote the values of integers encoded by $U$ and $V$ respectively. The corresponding representative $r(e_u, e_v)$ and deviation $d(e_u, e_v)$ can be computed based on the components of $U$ and $V$. The algorithm then stores $r(e_u, e_v)$ and $d(e_u, e_v)$ into $C_{t}$ at $C_{t}[e_u, e_v]$.

In addition to the tables we have described above, three $n \times \frac{n}{k}$ tables $D$, $E$, $G$ and one $\frac{n}{k} \times n$ table $F$ are needed to store the intermediate results of dynamic programming. $D$ is a table that stores the values of $M[i, j]$'s for all $i$ and $j$'s that are in the same aggregate interval. $D$ can be accessed by two integers. One is the starting position of a subsequence and the other one is the length of the subsequence. $E$ is a table that stores the values of $M[i, j]$'s for all $j$'s that are the first position of an aggregate interval. $E$ can be accessed by two integers. One is the starting postion of a subsequence and the other one is the integer identity of an aggregate interval, its first position is the ending position of the subsequence. $G$ is a table that stores the value of $M[i, j]$ for all $j$'s that are in the same aggregate interval. $G$ is accessed by the starting position of a subsequence and the relative position of the ending position of the subsequence in the aggregate interval that contains the ending position. $F$ is a table that stores the values of $M[i, j]$'s for all $i$'s that are the second position of an aggregate interval. Similarly, $F$ can also be accessed by two integers. One is the integer identity of an aggregate interval whose second position is the starting position of a subsequence and the other one is the ending position of the subsequence.

The algorithm computes the values of $M[i, j]$ in the same order as the one used in \cite{frid}. It follows the steps that are sketched below to compute the value of $M[i, j]$ for each pair of integers $i$ and $j$ that satisfy
$1 \leq i < j \leq n$.
\begin{enumerate}
\item{Precompute the central table $C_t$ as described above;}
\item{set $j$ to be $1$;}
\item{set $i$ to be $j-1$;}
\item{compute the value of $M[i, j]$ and store related values into the corresponding tables;}
\item{decrement $i$ by $1$;}
\item{go to step 4 if $i$ is larger than $0$;}
\item{increment $j$ by $1$;}
\item{go to step 3 if $j$ is not larger than $n$;}
\item{output $M[1, n]$ and the base pairs associated with $M[1, n]$.}
\end{enumerate}

We now describe the implementation of step 4 in detail. We consider three possible cases for integers $i$, $j$ and the computation of $M[i, j]$ is based on a different strategy for each case. The three cases are listed as follows.
\begin{enumerate}
\item{$i$ and $j$ are in the same aggregate interval $I_e$, where $0 \leq e \leq k$;}
\item{there exists integer $1 \leq e \leq k$ such that $i$ is in
aggregate interval $I_{e-1}$ and $j$ is in aggregate interval $I_{e}$;}
\item{there exists a set of aggregate intervals between $i$ and $j$.}
\end{enumerate}

For case 1, the algorithm computes the value of $M[i, j]$ with the same procedure as the one in the traditional dynamic programming algorithm. Specifically, the value of $M[i, j]$ can be computed by the following steps.
\begin{enumerate}
\item{Set $c_1$ to be $D[i, j-i]$;}
\item{set $c_2$ to be $D[i+1, j-i]$;}
\item{set $c_3$ to be $D[i+1, j-i-1]+C[S[i], S[j]]$;}
\item{exhaustively enumerate all integers $p$'s that
satisfy $i \leq p < j$ and set $c_4$ to be the maximum value of
$D[i, p-i+1]+D[p+1, j-p]$;}
\item{set $D[i, j-i+1]$ to be $\max{\{c_1, c_2, c_3, c_4\}}$}
\item{set $N[i]$ to be $D[i, j-i+1]$.}
\end{enumerate}
where $N$ is an array that stores the values of $M[i, j]$ for all $i$'s that satisfy $1 \leq i \leq j$. Similar to case 1, the computed value of $M[i, j]$ is also stored in $N[i]$ in cases 2 and 3.

For case 2, the algorithm computes the value of $M[i, j]$ with the same exhaustive search procedure as we have described for case 1. However, since $i$ and $j$ are in two different but consecutive aggregate intervals. $R_t$ needs to be queried to obtain the intermediate values needed for the exhaustive search. Based on the two aggregate intervals $I_{e-1}$ and $I_{e}$ that contain $i$ and $j$ respectively.  The steps for computing $M[i, j]$ are described as follows.
\begin{enumerate}
\item{Set $c_1$ to be $N[i+1]$;}
\item{set $x$ to be $R_{t}[j-1, e-1]$;}
\item{obtain the right difference vector associated with $x$ and
store its components in array $X$;}
\item{set $c_2$ to be $F[e-1, j-1]-\sum_{l=2}^{i-(e-1)w-1}{X(l)}$ if $i>(e-1)w+2$; set $c_2$ to be $F[e-1, j-1]$ if $i=(e-1)w+2$; set $c_2$ to be $F[e-1,j-1]+V(1)$ if $i=(e-1)w+1$;}
\item{set $c_3$ to be $c_2-V(i-(e-1)w)+C[S[i], S[j]]$;}
\item{set $v=R_{t}[j, e-1]$;}
\item{obtain the right difference vector associated with $v$ and
      store its components in array $V$;}
\item{exhaustively enumerate all $p$'s that satisfy $i \leq p \leq ew-1$ and set $c_4$ to be the maximum value of
$D[i, p-i+1]+F[e-1, j]-\sum_{l=2}^{p-(e-1)w}{V(l)}$;}
\item{if $i>(e-1)w+2$, exhaustively enumerate all $q$'s that satisfy $ew \leq q <j$ and set $c_5$ to be the maximum value of
$D[q+1, j-q]+F[e-1, q]-\sum_{l=2}^{i-(e-1)w-1}{V_{q}(l)}$, where $V_{q}$ stores the components of the right difference vector associated with $R_{t}[q,e-1]$;}
\item{if $i=(e-1)w+2$, exhaustively enumerate all $q$'s that satisfy $ew \leq q <j$ and set $c_5$ to be the maximum value of
$D[q+1, j-q]+F[e-1, q]$;}
\item{if $i=(e-1)w+1$, exhaustively enumerate all $q$'s that satisfy $ew \leq q <j$ and set $c_5$ to be the maximum value of
$D[q+1, j-q]+F[e-1, q]+V_{q}(1)$, where $V_{q}$ stores the components of the right difference vector associated with $R_{t}[q,e-1]$;}
\item{set $N[i]$ to be $\max{\{c_1, c_2, c_3, c_4, c_5\}}$;}
\item{set $G[i, j-ew]$ to be $N[i]$.}
\end{enumerate}

For case 3, there exists a few consecutive aggregate intervals between $i$ and $j$, the computation of $M[i, j]$ needs to query both $L_t$ and $R_t$ to efficiently obtain the $(i, j)$ representative and deviation for each such aggregate interval. Specifically, we assume $i$, $j$ are contained in aggregate intervals $I_{m}$ and $I_{n}$ respectively. The steps that compute $M[i, j]$ are described as follows.
\begin{enumerate}
\item{Set $c_1$ to be $N[i+1]$;}
\item{set $x$ to be $R_{t}[j-1, m]$;}
\item{obtain the right difference vector associated with $x$ and
store its components in array $X$;}
\item{set $c_2$ to be $F[m, j-1]-\sum_{l=2}^{i-mw-1}{X(l)}$ if $i>mw+2$; set $c_2$ to be $F[m, j-1]$ if $i=mw+2$; set $c_2$ to be $F[m,j-1]+V(1)$ if $i=mw+1$;}
\item{set $c_3$ to be $c_2-V(i-mw)+C[S[i], S[j]]$;}
\item{set $v=R_{t}[j, m]$;}
\item{obtain the right difference vector associated with $v$ and
      store its components in array $V$;}
\item{exhaustively enumerate all $p$'s that satisfy $i \leq p \leq (m+1)w-1$ and set $c_4$ to be the maximum value of
$D[i, p-i+1]+F[m, j]-\sum_{l=2}^{p-mw}{V(l)}$;}
\item{if $i>mw+2$, exhaustively enumerate all $q$'s that satisfy $ew \leq q <j$ and set $c_5$ to be the maximum value of
$D[q+1, j-q]+F[m, q]-\sum_{l=2}^{i-mw-1}{V_{q}(l)}$, where $V_{q}$ stores the components of the right difference vector associated with $R_{t}[q,m]$;}
\item{if $i=mw+2$, exhaustively enumerate all $q$'s that satisfy $ew \leq q <j$ and set $c_5$ to be the maximum value of
$D[q+1, j-q]+F[m, q]$;}
\item{if $i=mw+1$, exhaustively enumerate all $q$'s that satisfy $ew \leq q <j$ and set $c_5$ to be the maximum value of
$D[q+1, j-q]+F[e-1, q]+V_{q}(1)$, where $V_{q}$ stores the components of the right difference vector associated with $R_{t}[q,m]$;}
\item{exhaustively enumerate all $f$'s that satisfy $m+1 \leq f \leq n-1$ and execute steps 13 to 17 for each $f$;}
\item{set $x$ to be $R_{t}[j, f]$;}
\item{set $y$ to be $L_{t}[i, f]$;}
\item{query $C_t$ with $x$, $y$ and store the deviation stored in $C_t[x, y]$ into $d$;}
\item{store the value of $d+E[i, f]+F[f, j]$ into an array $H$;}
\item{set $c_6$ to be the largest number in $H$;}
\item{set $N[i]$ to be $\max{\{c_1, c_2, c_3, c_4, c_5, c_6\}}$;}
\item{set $G[i,j-nw]$ to be $N[i]$.}
\end{enumerate}

After the values of all $M[i, j]$'s are computed, the algorithm computes the $(j, s)$ right difference vectors for each $s$ that satisfies $(s+1)w<j$ and write the integer encoded by it into $R_t[j, s]$. The computation of all these right different vectors can be based on array $N$. In addition, when $j=gw$ for some integer $g$, the algorithm computes the $(i, g)$ left difference vector for each $i \leq (g-1)w$ based on $G$ and stores the integer encoded by it into $L_t[i, g]$. Let $w=\frac{1}{4}\log_{2}{n}$, we immediately obtain the following theorem.

\begin{theorem}
\rm
\label{th1}
Given an RNA sequence $S$ of length $n$, the algorithm correctly computes the value of $M[i, j]$ for all integers $i$ and $j$ that satisfy $1 \leq i \leq j \leq n$ in time $O(\frac{n^3}{\log_{2}{n}})$ and space $O(\frac{n^2}{\log_{2}{n}})$.

\begin{proof}
It is straightforward to see that the algorithm correctly computes $M[i, j]$ in cases 1 and 2 since an exaustive search among all integers between $i$ and $j$ is performed in both cases to compute $M[i, j]$. In case 3, the algorithm queries $C_t$ with the integers encoded by the $(i, s)$ left difference vector and $(j, s)$ right difference vector to compute the value of $(i, j)$ deviation for each aggregate interval $I_s$ between $i$ and $j$. From Lemma \ref{lm1} and Lemma \ref{lm2}, the value of $M[i, j]$ can be correctly computed based on the deviations obtained by querying $C_t$. In addition, the algorithm correctly updates the content of $L_t$ and $R_t$ after all $M[i, j]$'s have been computed for each $j$.  The correctness of the algorithm thus follows.

We then analyze the time complexity of the algorithm. The precomputation stage computes the central table $C_t$. $C_t$ contains $2^{2w}$ locations in total and the number in each location can be computed in time $O(w)$. The precomputation stage thus needs at most $O(2^{2w}w)$ time. Next, we consider the total number of pairs $(i, j)$ for which $M[i, j]$ needs to be computed, where $1 \leq i \leq j \leq n$. It is not difficult to see that $O(nw)$ of these pairs are processed as case 1 or 2 and the remaining pairs are processed as case 3. For each pair that is processed as case 3, $C_t$ needs to be queried for $O(\frac{n}{w})$ times and the time needed to compute the value of $M[i, j]$ is thus $O(\frac{n}{w})$. In addition, the content of $R_t$ can be updated in time $O(n)$ for a fixed $j$. Similarly, the content of $L_t$ can be updated in time $O(nw)$ and it needs to be updated for up to $O(\frac{n}{w})$ times in total. The total computation time $T(n)$ of the algorithm thus can be bounded from above as follows.
\begin{equation}
T(n) \leq A(2^{2w}w+nw^{3}+\frac{n^3}{w}+n^{2})
\end{equation}
where $A$ is some positive constant. The algorithm thus needs $O(\frac{n^3}{\log_{2}{n}})$ time if $w=\frac{1}{4}\log_{2}{n}$.

The amount of memory space needed by $R_t$, $L_t$, $D$, $E$, $G$, $F$ are $O(\frac{n^{2}}{w})$. $C_t$ can be implemented with $O(2^{2w})$ space. $N$ needs $O(n)$ space and can be reused for each $j$. The total amount of space $S(n)$ needed by the algorithm can thus be bounded from above as follows.
\begin{equation}
S(n) \leq B(\frac{n^{2}}{w}+2^{2w}+n)
\end{equation}
where $B$ is some positive constant and the space complexity of the algorithm is $O(\frac{n^2}{\log_{2}{n}})$ if $w=\frac{1}{4}\log_{2}{n}$.               
\end{proof}
\end{theorem}

The algorithm needs to implement a trace-back procedure to obtain the base pairs in the predicted secondary structure of the sequence. However, the trace-back procedure of the $O(n^3)$ traditional dynamic programming algorithm cannot be directly used since the algorithm does not maintain an $n \times n$ table that stores the information needed for tracing back. We show that the trace-back procedure can be performed with a recursive algorithm in time $O(n^2)$ and $O(n)$ additional space. We use TRACEBACK to name this algorithm and it needs two parameters $i$ and $j$, where $1 \leq i \leq j \leq n$. It returns the base pairs contained in the secondary structure that corresponds to $M[i, j]$ in subsequence $S[i \cdots j]$. For the convenience of notation, we use $T_{r}(i,j)$ to denote the set of base pairs returned by TRACEBACK on subsequence $S[i \cdots j]$. The steps of TRACEBACK can be described as follows.
\begin{enumerate}
\item{Return an empty set if $i>j$ or $i=j$;}
\item{obtain the values of $M[i, j]$, $M[i, j-1]$, $M[i+1, j]$, and $M[i+1, j-1]$ and store them into $c$, $c_1$, $c_2$, and $c_3$ by querying tables $R_t$ and $F$;}
\item{obtain the values of $M[i, p]$ and $M[p+1, j]$ for each $p$ that satisfies $i \leq p < j$ and store their values into arrays $O$ and $Q$ by querying $L_t$, $R_t$, $E$, $F$ and $D$;}
\item{if $c=c_1$, recursively call TRACEBACK on $i$ and $j-1$ and return $T_{r}(i, j-1)$;}
\item{if $c=c_2$, recursively call TRACEBACK on $i+1$ and $j$ and return $T_{r}(i+1, j)$;}
\item{set $c_3$ to be $c_3+C[S[i], S[j]]$}
\item{if $c=c_3$, recursively call TRACEBACK on $i+1$ and $j-1$ and return $\{(i, j)\}\cup T_{r}(i+1, j-1)$;}
\item{for each $p$ that satisfies $i \leq p < j$, find the integer $p$ that satisfies $c=O[p]+Q[p+1]$;}
\item{recursively call TRACEBACK on $i$ and $p$ and store the returned result in $S_l$;}
\item{recursively call TRACEBACK on $p+1$ and $j$ and store the returned result in $S_r$;}
\item{return $S_l \cup S_r$.}
\end{enumerate}

\begin{theorem}
\rm
\label{th2}
Algorithm TRACEBACK returns the base pairs in the predicted secondary structure of $S$ in time $O(n^2)$ and needs $O(n)$ additional space.

\begin{proof}
The correctness of the algorithm follows directly from Theorem \ref{th1}. We use $T(r)$ to denote the computation time needed by TRACEBACK on a subsequence of length $r$. Since the computation of array $O$, $Q$ and integer $p$ can be performed in $O(r)$ time, the following inequality holds for $T(r)$.
\begin{equation}
T(r) \leq T(r_1)+T(r-r_1)+Ar
\end{equation}
where $r_1$ is an integer that satifies $1 \leq r_1 \leq r-1$ and $A$ is some positive constant.

We define a positive constant $A_0$ as follows.
\begin{equation}
A_0=\max{\{\max_{1 \leq r_1 \leq 4}{\{\frac{T(r_1)+r_1}{{r_{1}}^{2}}\}}, 4A\}}
\end{equation}
We then show $T(r) \leq A_{0}r^{2}-r$ for all $r>0$ by induction. First, for $r_0 \leq 4$, the inequality holds from the definition of $A_0$. We now assume that the inequality holds for all integers that are at most $r$, where $r \geq 4$. For $T(r+1)$, we have
\begin{eqnarray}
T(r+1) & \leq & T(r_1)+T(r+1-r_1)+Ar \\
       & \leq & A_{0}((r+1)^{2}-2r_{1}(r+1-r_{1}))-(r+1)+A(r+1) \\
       & = & (A_{0}(r+1)^{2}-(r+1))-(2A_{0}r_{1}(r+1-r_{1})-A(r+1)) \\
       & \leq & A_{0}(r+1)^{2}-(r+1)
\end{eqnarray}
where the second inequality is due to the assumption of induction and the last inequality comes from the fact that $r$ is at least $4$ and $A_0 \geq 4A$. The inequality thus holds for all $r>0$ from the principle of induction. The inequality suggests that TRACEBACK needs $O(n^2)$ time to determine all base pairs in the predicted secondary structure of $S$.

It is not difficult to see that the space of array $O$ and $Q$ can be reused when TRACEBACK calls itself recursively. TRACEBACK therefore only needs $O(n)$ additional space for its computation.
\end{proof}
\end{theorem}

Putting Theorems \ref{th1} and \ref{th2} together, we immediately obtain the following theorem on the time and space complexities of the basic RNA folding problem.

\begin{theorem}
\rm
\label{th3}
The basic RNA folding problem can be solved in time $O(\frac{n^3}{\log_{2}{n}})$ while using only $O(\frac{n^{2}}{\log_{2}{n}})$ memory space, where $n$ is the length of the sequence.
\end{theorem}

It is worth pointing out that the above result for space complexity does not rely on the word width assumption. To see this, we assume the word width assumption does not hold and an integer between $0$ and $n$ may need $O(\log_{2}n)$ words for storage. If this is the case, the algorithm can create $n+1$ blocks of memory space to store the integers from $0$ to $n$. For a location in any table used by the algorithm, a pointer is stored in the location. This pointer points to the block that stores the value of the integer that is originally stored directly in the location. It can be seen that the above strategy does not increase the amount of memory needed by the tables. Since the blocks used to store integers from $0$ to $n$ need only $O(n\log_{2}n)$ space, the space complexity of the algorithm remains to be $O(\frac{n^{2}}{\log_{2}{n}})$. However, an additional $O(\log_{2}{n})$ factor needs to be added to the time complexity of the algorithm since one step for integer arithmetic may need up to $O(\log_{2}{n})$ time. We therefore obtain the following corollary.

\begin{corollary}
\rm
The basic RNA folding problem can be solved in time $O(n^{3})$ while using only $O(\frac{n^{2}}{\log_{2}{n}})$ space when the word width assumption does not hold, where $n$ is the length of the sequence.
\end{corollary}

\section{Improve the Time Complexity to $O(\frac{n^{3}}{\log_{2}^{2}{n}})$}
\label{sec1}

In this section, we develop strategies to further improve the time complexity of the algorithm we have developed in the previous section. Specifically, the algorithm does not compute the value of $M[i, j]$ for each individual pair $(i, j)$. Instead, for a fixed $j$, strategies based on the Four-Russians technique are used to determine the values of $M[i, j]$ for all $i$'s that are in the same aggregate interval in $O(1)$ time. In other words, the values of $M[i, j]$'s are determined in a blockwise fashion for each fixed $j$ and the computation time needed by the dynamic programming is thus further reduced to $O(\frac{n^{3}}{\log_{2}^{2}{n}})$. However, additional memory space is needed to store the results of precomputation and the space complexity of the algorithm becomes $O(n^{2.5})$.

For each pair $(i, j)$, the algorithm we have developed in the previous section fully determines the value of $M[i, j]$ before it starts computing $M[i-1, j]$. To improve the time complexity of the algorithm, we observe that it does not have to follow this order of evaluation to compute the values of all $M[i, j]$'s. Instead, for a fixed $j$, the algorithm can process all aggregate intervals to the left of $j$ in an order from right to left. A partial optimal result can be computed for each pair of $(i, j)$ when an aggregate interval is processed. The values of all $M[i, j]$'s are finally determined when all aggregate intervals to the left of $j$ are processed.

\begin{definition}
\rm
\label{def1}
Let $S$ be an RNA sequence of length $n$. $i$, $j$, $t$ and $o$ are integers that satisfy $1 \leq i \leq t \leq o < j \leq n$,  a $(t, o)$ {\it partial optimal result} $P[i, j, t, o]$ for $(i, j)$ is defined to be
\begin{equation}
P[i, j, t, o]=\max_{t \leq l \leq o}{\{M[i, l]+M[l+1,j]\}}
\end{equation}
\end{definition}

\begin{definition}
\rm
\label{def2}
Let $S$ be an RNA sequence of length $n$. $i$, $j$, $t$ and $o$ are integers that satisfy $1 \leq i \leq t \leq o < j \leq n$, a set of non-crossing base pairs in subsequence $S[i \cdots j]$ is a $(t, o)$ {\it partial set} for $(i, j)$ if it does not contains a base pair formed between $S[g]$ and $S[h]$, where $i \leq g<t$ and $o < h \leq j$. A $(t, o)$ partial set for $(i, j)$ is {\it optimal} if the number of base pairs in the set is the maximum of all $(t, o)$ partial sets for $(i, j)$.
\end{definition}

\begin{lemma}
\rm
\label{lm3}
Let $S$ be an RNA sequence of length $n$. $i$, $j$, $t$ and $o$ are integers that satisfy $1 \leq i \leq t \leq o < j \leq n$. $P[i, j, t, o]$
is the number of base pairs in an optimal $(t, o)$ partial set for $(i, j)$ may contain.

\begin{proof}
We use $O_t$ to denote the set of all $(t, o)$ partial sets for $(i, j)$, $L$ is the optimal $(t, o)$ partial set in $O_t$ and $N_l$ is the number of base pairs $L$ contains. We first show that $P[i, j, t] \geq N_l$.

We consider the nucleotides in positions in the subsequence $S[t \cdots o]$. These nucleotides are {\it middle nucleotides}. If none of the base pairs in $L$ contain a middle nucleotide, we can partition the base pairs in $L$ into two disjoint subsets $L_l$ and $L_r$, where $L_l$ contains the base pairs formed among nucleotides in subsequence $S[i \cdots t]$ while $L_r$ contains those formed among nucleotides in subsequence $S[t+1 \cdots j]$. We thus have
\begin{eqnarray}
N_l & \leq & M[i, t]+M[t+1, j] \\
    &  \leq & P[i, j, t, o]
\end{eqnarray}
where the first inequality is due to the fact that $M[i, t]$ and $M[t+1, j]$ are the maximum number of non-crossing base pairs that can be formed in subsequences $S[i \cdots i]$ and $S[t+1 \cdots j]$ respectively; the second inequality comes from Definition \ref{def1}.

If at least one base pair in $L$ contains a middle nucleotide. Without loss of generality, we assume there exists a base pair formed between a nucleotide in position $b_{p}$, where $i \leq b_{p}< t$, and a middle nucleotide. Such a base pair is a {\it left base pair}. We find all middle nucleotides contained in left base pairs in $L$ and select the rightmost one. We use $r_t$ to denote its position. Similarly, base pairs in $L$ can be partitioned into two disjoint subsets $L_l$ and $L_r$ where $L_l$ contains the base pairs formed among nucleotides in subsequence $S[i \cdots r_t]$ and $L_r$ contains those formed among nucleotides in subsequence $S[r_t+1 \cdots j]$. We thus have
\begin{eqnarray}
N_l & \leq & M[i, r_t]+M[r_t+1, j] \\
    & \leq &  P[i, j, t, o]
\end{eqnarray}
where the second inequality is due to the fact that $t \leq r_t \leq o$. $N_l \leq P[i, j, t, o]$ thus holds.

We then show that $P[i, j, t] \leq N_l$. From Definition \ref{def1}, there exists $l_{t}$ such that $P[i, j, t, o]=M[i, l_t]+M[l_{t}+1, j]$ and $t \leq l_{t} \leq o$. $M[i, l_{t}]$ and $M[l_{t}+1, j]$ correspond to two sets of base pairs in subsequences $S[i \cdots l_{t}]$ and $S[l_{t}+1 \cdots j]$ respectively. We use $S_{i, l_{t}}$ and $S_{l_{t}+1, j}$ to denote them. It is clear that $S_{i, l_{t}}$ and $S_{l_{t}+1, j}$ are disjoint and $S_{i, l_{t}} \cup S_{l_{t}+1, j} \in O_{t}$. We thus obtain
\begin{eqnarray}
    P[i, j, t, o] & = & |S_{i, l_t}|+|S_{l_{t}+1, j}| \\
               & \leq & N_l
\end{eqnarray}
where $|S_{i, l_{t}}|$ and $|S_{l_{t}+1, j}|$ are the number of base pairs in $S_{i, l_{t}}$ and $S_{l_{t}+1, j}$ respectively. The inequality is due to the fact that $L$ is optimal. $N_l=P[i, j, t, o]$ thus follows.
\end{proof}
\end{lemma}

It is not difficult to see that Definitions \ref{def1} and \ref{def2} provide an approach to computing partial optimal results for $M[i, j]$ during the dynamic programming procedure. From Lemma \ref{lm3}, it is also clear that each $(t, o)$ partial optimal result for $(i, j)$ corresponds to an optimal $(t, o)$ partial set for $(i, j)$ in subsequence $S[i \cdots j]$. The following lemma shows that a property similar to the one shown in Lemma \ref{lm1} for $M[i, j]$ also holds for $P[i, j, t, o]$.

\begin{lemma}
\rm
\label{lm4}
Let $i$, $j$, $t$, and $o$ be integers that satisfy $i \leq t \leq o < j$,  we have $P[i, j, t, o] \leq P[i-1, j, t, o] \leq P[i, j, t, o]+1$ for $P[i, j, t, o]$ and $P[i-1, j, t, o]$.

\begin{proof}
From Lemma \ref{lm3}, there exists an optimal $(t, o)$ partial set $G_{t}$ for $(i-1, j)$ such that $G_t$ contains $P[i-1, j, t, o]$ non-crossing base pairs. If $S[i]$ is in a base pair in $L_{t}$, we use $p_{i}$ to denote this base pair. It is not difficult to see that $G_{t}-\{p_{i}\}$ is a $(t, o)$ partial set for $(i, j)$. From Lemma \ref{lm3}, we obtain
\begin{equation}
|G_{t}|-1 \leq P[i, j, t, o]
\end{equation}
where $|G_{t}|$ is the number of stems in $G_t$. Since $|G_t|=P[i-1, j, t, o]$, we immediately obtain
\begin{equation}
P[i-1, j, t, o] \leq P[i, j, t, o]+1
\end{equation}

If $S[i]$ is not contained in any base pair in $G_{t}$, $G_{t}$ is a $(t, o)$ partial set for $(i, j)$. From Lemma \ref{lm3}, we obtain
\begin{equation}
|G_{t}| \leq P[i, j, t, o]
\end{equation}
Similarly, since $|G_(t)|=P[i-1, j, t, o]$, we immediately obtain
\begin{equation}
P[i-1, j, t, o] \leq P[i, j, t, o]
\end{equation}
Putting two possible cases together, we have
\begin{equation}
P[i-1, j, t, o] \leq P[i, j, t, o]+1
\end{equation}

On the other hand, there exists an optimal $t$ partial set $G'_{t}$ for $(i, j)$ such that $G'_{t}$ contains $P[i, j, t, o]$ non-crossing stems. It is straightforward to see that $G'_t$ is also a $(t, o)$ partial set for $(i-1, j)$. We thus can obtain
\begin{equation}
|G'_{t}| \leq P[i-1, j, t, o]
\end{equation}
Since $|G'_{t}|=P[i, j, t, o]$, we have $P[i, j, t, o]\leq P[i-1, j, t, o]$. The lemma thus follows.
\end{proof}
\end{lemma}

Lemma \ref{lm4} provides the foundation of our algorithm. Specifically, for an aggregate interval $I_s$, where $0 \leq s \leq k-1$, and fixed integers $j$ and $t$ such that $(s+1)w < t < j$. The values of $P[(s+1)w, j, t, j-1], P[(s+1)w-1, j, t, j-1], \cdots, P[sw+1, j, t, j-1]$ can be represented by $P[sw+1, j, t, j-1]$ and a $w$-dimensional binary vector $V_r$. Such a binary vector is a {\it $(j, s, t)$ partial right difference vector}.

In addition, consider two aggregate intervals $I_{s_1}$, $I_{s_2}$ and a fixed integer $j$ such that $s_1 < s_2$ and $(s_2+1)w < j$. The values of $P[(s_{1}+1)w, j, s_{2}w+1, (s_{2}+1)w], P[(s_{1}+1)w-1, j, s_{2}w+1, (s_{2}+1)w], \cdots, P[s_{1}w+1, j, s_{2}w+1, (s_{2}+1)w]$ can be represented by $P[s_{1}w+1, j, s_{2}w+1, (s_{2}+1)w]$ and a $w$-dimensional binary vector $V_c$. Such a binary vector is a {\it $(j, s_1, s_2)$ partial central difference vector}.

\begin{definition}
\rm
Let $S$ be an RNA sequence of length $n$. $i$, $j$, $t$ and $o$ are integers that satisfy $1 \leq i \leq t \leq o < j$. The {\it $(t, o)$ partial deviation} for $(i, j)$ is defined to be
$P[i, j, t, o]-M[i, o]-M[o+1, j]$.
\end{definition}

\begin{definition}
\rm
Let $S$ be an RNA sequence $S$ of length $n$. A {\it partial right table} $R_{p}$ is an $n \times (k-1)$ table such that $R_{p}[j, s]$ stores the integer encoded by the $(j, s, t)$ partial right difference vector for some integer $t$ that satisfies $(s+1)w < t < j$ and the corresponding $(t, j-1)$ partial deviation for $((s+1)w, j)$.
\end{definition}

\begin{definition}
\rm
Let $S$ be an RNA sequence of length $n$, $I_0, I_1, \cdots, I_{k-1}$ are the aggregate intervals of length $w$ in $S$. A {\it central partial table} $C_{p}$ is a $k \times k \times 2^{w+1}$ table such that $C_{p}$ can be accessed with three integers $s_1$, $s_2$ and $v_c$ that satisfy $0 \leq s_1 < s_2 \leq k$ and $0 \leq v_c < 2^{w+1}$. $C_p[s_1, s_2, v_c]$ stores two positive integers whose values are at most $w$.
\end{definition}

\begin{lemma}
\rm
\label{lm5}
Let $S$ be an RNA sequence of length $n$, $I_0, I_1, \cdots, I_{k-1}$ are the aggregate intervals of length $w$ in $S$. Given integers $s_1$, $s_2$ and $j$ that satisfy $0 \leq s_1 < s_2 \leq k-1$ and $(s_2+1)w<j$. The $(j, s_1, s_2)$ partial central difference vector and the value of $(s_{2}w+1, (s_{2}+1)w)$ partial central deviation for $((s_{1}+1)w, j)$ can be computed in time $O(w^3)$ if the following conditions hold.
\begin{enumerate}
\item{The $(j, s_2)$ right difference vector is available;}
\item{for each $j_{s}$ that satisfies $s_{2}w+1 \leq j_{s} \leq (s_{2}+1)w$, the $(j_{s}, s_1)$ right difference vector is available;}
\item{the $((s_1+1)w, s_2)$ left difference vector is available;}
\end{enumerate}

\begin{proof}
We use $U$, $X$ to denote the $(j, s_2)$ right difference vector and the $((s_1+1)w, s_2)$ left difference vector respectively. In addition, $V_1, V_2, \cdots, V_{w}$ denote the right difference vectors for all positions in $I_{s_2}$. Specifically, for $1 \leq m \leq w$, $V_{m}$ is the $(s_{2}w+m, s_1)$ right difference vector. We show how the $(j, s_1, s_2)$ partial central difference vector can be computed from $U$, $X$ and $V_1, V_2, \cdots, V_{m}$.

We assume $W=M[(s_1+1)w, (s_2+1)w]+M[(s_2+1)w+1, j]$. Given two integers $s_b$, $s_c$ such that $1 \leq s_{b} \leq w$ and $1 \leq s_{c} \leq w$. We define $J(s_b, s_c)$ as follows.
\begin{equation}
J(s_b, s_c)=M[s_{1}w+s_b, s_{2}w+s_c]+M[s_{2}w+s_c+1, j]
\end{equation}
It is not difficult to see that $J(s_b, s_c)$ can be computed from $W$ and $U$, $X$ and $V_1, V_2, \cdots, V_m$ as follows.
\begin{eqnarray}
J(s_b, s_c)& = & M[(s_1+1)w, s_{2}w+s_c]+\sum_{l=s_{b}}^{w-1}{V_{s_c}(l)}+M[s_{2}w+s_c+1, j] \\
           & = & M[(s_1+1)w, s_{2}w+s_c]+M[(s_2+1)w+1, j]+
                 \sum_{l=s_{b}}^{w-1}{V_{s_c}(l)}+
                 \sum_{l=s_{c}+1}^{w}{U(l)} \\
           & = & W-\sum_{l=s_c+1}^{w}{X(l)}+
                  \sum_{l=s_b}^{w-1}{V_{s_c}(l)}+
                  \sum_{l=s_{c}+1}^{w}{U(l)}
\end{eqnarray}

From Definition \ref{def1}, $P[s_{1}w+s_{b}, j, s_{2}w+1, (s_{2}+1)w]$ can be computed from $J(s_b, s_c)$ as follows.
\begin{equation}
P[s_{1}w+s_{b}, j, s_{2}w+1, (s_{2}+1)w]=\max_{1 \leq s_c \leq w}{\{J(s_b, s_c)\}}
\end{equation}
The computation of $J(s_b, s_c)$ needs $O(w)$ time and there are in total $O(w^2)$ pairs of $(s_b, s_c)$'s. The $(j, s_1, s_2)$ partial central difference vector and the value of $(s_{2}w+1, (s_{2}+1)w)$ partial deviation for $((s_{1}+1)w, j)$ can thus be computed in $O(w^3)$ time.
\end{proof}
\end{lemma}

From Lemma \ref{lm5}, the content of $C_{p}$ can be precomputed for each combination of a pair of two different aggregate intervals $I_{s_1}$, $I_{s_2}$ that are both to the left of $j$ and a $w$-dimensional binary vector $X$ that represents the
$(j, s_2)$ right difference vector. The integer encoded by the corresponding partial central difference vector and the value of the corresponding partial central deviation can be stored in $C_p$. After the precomputation of $C_p$ is performed. The algorithm can query $C_p$ to obtain the integer encoded by the $(j, s_1, s_2)$ partial central difference vector and the value of $(s_{2}w+1, (s_{2}+1)w)$ partial central deviation for $((s_{1}+1)w, j)$ in $O(1)$ time. However, the precomputation of $C_p$ must be interleaved with the computation of $M[i, j]$'s during the dynamic programming procedure.

\begin{definition}
\rm
Let $S$ be an RNA sequence $S$ of length $n$. An {\it updation  table} $U_{p}$ is a four-dimensional table and two integers are stored in a location $U_{p}[d_1, d_2, u_1, u_2]$, where $d_1$, $d_2$, $u_1$ and $u_2$ satisfy $0 \leq d_1 \leq n$, $0 \leq d_2 \leq 2n$, $0 \leq u_1 < 2^{w+1}$ and $0 \leq u_2 < 2^{w+1}$. One of the integers stored in a location in $U_p$ is at most $2^{w+1}-1$ and the other one is at most $n$.
\end{definition}

\begin{lemma}
\rm
\label{lm6}
Let $S$ be an RNA sequence of length $n$, $I_0, I_1, \cdots, I_{k-1}$ are the aggregate intervals of length $w$ in $S$. Given integers $s_1$, $s_2$ and $j$ that satisfy $0 \leq s_1 < s_2 \leq k-1$ and $(s_2+1)w<j$, the $(j, s_1, s_{2}w+1)$ partial right difference vector and the value of the $(s_{2}w+1, j-1)$ partial deviation for $((s_{1}+1)w, j)$ can be computed in time $O(w)$ if the following conditions hold.
\begin{enumerate}
\item{The $(j, s_2, (s_{2}+1)w+1)$ partial right difference vector is available;}
\item{the $((s_{2}+1)w+1, j-1)$ partial deviation for
$((s_{1}+1)w, j)$ is available;}
\item{the $(j, s_1, s_2)$ partial central difference vector is
available;}
\item{the $(s_{2}w+1, (s_{2}+1)w)$ partial deviation for
$((s_{1}+1)w, j)$ is available;}
\item{the values of $M[(s_{1}+1)w, j-1]$, $M[(s_{1}+1)w, (s_{2}+1)w]$, and $M[(s_{2}+1)w+1, j]$ are available.}
\end{enumerate}

\begin{proof}
We use $U$ to denote the $(j, s_2, (s_{2}+1)w+1)$ partial right difference vector. $d_{l}$ denotes the $((s_{2}+1)w+1, j-1)$ partial deviation for $((s_{1}+1)w, j)$; $V$ denotes the
$(j, s_1, s_2)$ partial central difference vector. $d_{r}$ denotes the $(s_{2}w+1, (s_{2}+1)w)$ partial deviation for
$((s_{1}+1)w, j)$.

For an integer $i_s$ such that $1 \leq i_s \leq w$, we use
$A_{1}[i_s]$ to denote $P[s_{1}w+i_s, j, (s_2+1)w+1, j-1]$ and
$A_{2}[i_s]$ to denote $P[s_{1}w+i_s, j, (s_{2}w+1, (s_2+1)w]$. It is not difficult to see that the values of $A_{1}[i_s]$ and
$A_{2}[i_s]$ can be computed based on $U$, $V$, $d_1$ and $d_2$ as follows.
\begin{equation}
A_{1}[i_s]=M[(s_{1}+1)w, j-1]+d_{l}+\sum_{l=i_s}^{w-1}{U(l)}
\end{equation}
\begin{equation}
A_{2}[i_s]=M[(s_{l}+1)w, (s_{2}+1)w]+M[(s_{2}+1)w+1, j]+d_{r}+\sum_{l=i_s}^{w-1}{V(l)}
\end{equation}

Since the values of $M[(s_{1}+1)w, j-1]$, $M[(s_{1}+1)w, (s_{2}+1)w]$, and $M[(s_{2}+1)w+1, j]$ are available, we are able to compute the values of $A_{1}[i_s]$ and $A_{2}[i_s]$. It is straightforward to see that the following equality holds.
\begin{equation}
P[(s_{1}w+i_s, j, s_{2}w+1, j-1]=\max{\{A_{1}[i_s], A_{2}[i_s]\}}
\end{equation}
The $(j, s_1, s_{2}w+1)$ partial right difference vector and the value of the $(s_{2}w+1, j-1)$ partial deviation for $((s_{1}+1)w, j)$ can thus be computed in time $O(w)$.
\end{proof}
\end{lemma}

From Lemma \ref{lm6}, the content of $U_{p}$ can be precomputed for each possible combinations of $U$, $V$, $d_{l}$, and
$M[(s_{l}+1)w, (s_{2}+1)w]+M[(s_{2}+1)w+1, j]+d_{r}-M[(s_{1}+1)w, j-1]$. For each such combination, the corresponding partial right difference vector and partial deviation can be computed and stored in $U_{p}$. Specifically, for each such combination, we use $e_u$, $e_v$ to denote the integers encoded by the binary vectors in $U$ and $V$ and let
\begin{equation}
d_1=d_{l}
\end{equation}
\begin{equation}
d_2=M[(s_{l}+1)w, (s_{2}+1)w]+M[(s_{2}+1)w+1, j]+d_{2}-M[(s_{1}+1)w, j-1]+n
\end{equation}
It is clear that $0 \leq d_1 \leq n$ and $0 \leq d_2 \leq 2n$. The integer encoded by the computed partial right difference vector and partial deviation are then stored in $U_{p}[d_1, d_2, e_u, e_v]$. The precomputation of $U_p$ is performed before the dynamic programming procedure starts.

\begin{lemma}
\rm
\label{lm7}
Let $S$ be an RNA sequence of length $n$, $I_0, I_1, \cdots, I_{k-1}$ are the aggregate intervals of length $w$ in $S$. Given integers $s$ and $j$ that satisfy $0 \leq s \leq k-1$ and $(s+1)w<j$, the $(j, s)$ right difference vector can be computed in time $O(w^{2})$ if the following conditions hold.
\begin{enumerate}
\item{The $(j, s, (s+1)w+1)$ right partial difference vector is available;}
\item{the $((s+1)w+1, j-1)$ partial value for $((s+1)w, j)$ is available;}
\item{the value of $M[i, j]$ is available for each $i$ that satisfies $(s+1)w+1 \leq i < j$ and its value can be obtained in $O(1)$ time;}
\item{the value of $M[i, j_{t}]$ is available for each $j_t$ that satisfies $1 \leq i < j_{t} < j$ and its value can be obtained in $O(1)$ time.}
\end{enumerate}

\begin{proof}
We use $U$ to denote the $(j, s, (s+1)w+1)$ right partial difference vector. $d_{p}$ to denote the $((s+1)w+1, j-1)$ partial value for $((s+1)w, j)$; The values of $M[i, j]$ for each $i$ in $I_s$ can be computed with the following steps.
\begin{enumerate}
\item{Set $i$ to be $(s+1)w$;}
\item{exhaustively enumerate each integer $q$ that satisfies
$i \leq q \leq (s+1)w$ and compute the maximum value of
$M[i, q]+M[q+1, j]$ and set $c_1$ to be this maximum value;}
\item{set $c_2$ to be $M[i, j-1]+d_l+\sum_{l=i-sw}^{w-1}{U(l)}$;}
\item{set $c_3$ to be $M[i-1, j-1]+C[S[i], S[j]]$;}
\item{set $c_4$ to be $M[i-1, j]$;}
\item{set $c_5$ to be $M[i, j-1]$;}
\item{set $M[i, j]$ to be $\max{\{c_1, c_2, c_3, c_4, c_5\}}$ and store the result appropriately in memory;}
\item{decrement $i$ by 1;}
\item{if $i>sw$, go to step 2;}
\item{compute the $(j, s)$ right difference vector based on the values of $M[i, j]$'s that have been computed for $sw+1 \leq i \leq (s+1)w$.}
\end{enumerate}

It is clear that the above algorithm correctly computes the values of $M[i, j]$'s for all $i$'s that satisfy $sw+1 \leq i \leq (s+1)w$. Since the time needed to compute $M[i, j]$ for a single $i$ is $O(w)$, the $(j, s)$ right difference vector can be computed in time $O(w^2)$. The lemma thus follows.
\end{proof}
\end{lemma}

In addition to tables $L_t$, $R_t$ that have been defined in Section \ref{sec0} and all tables defined in this section, the algorithm uses an $n \times n$ table $M_r$ to store the value of $M[i, j]$ for each $i$ and $j$ that satisfy $1 \leq i \leq j \leq n$. The steps of the algorithm can be sketched as follows.
\begin{enumerate}
\item{Precompute table $U_p$ as described above;}
\item{set $j$ to be $1$;}
\item{set $s$ to be $\lfloor \frac{j}{w} \rfloor$;}
\item{for each value $i$ in $I_{s}$ and $i<j$, compute $M[i, j]$ and store its value in $M_r[i, j]$;}
\item{for each value $i$ in $I_0, I_1, \cdots, I_{s-1}$, compute
$P[i, j, sw+1, j-1]$;}
\item{for each $h$ that satisfies $0 \leq h \leq s-1$, compute the $(j, h, sw+1)$ partial right difference vector $U_h$ and the $(sw+1, j-1)$ partial deviation $d_h$ for $((h+1)w, j)$, store the integer encoded by $U_h$ and $d_h$ into $R_{p}[j, h]$;   }
\item{set $y$ to be $s-1$;}
\item{compute the $(j, y)$ right difference vector $V_t$ with the algorithm in Lemma \ref{lm7} and store the integer encoded by $V_t$ into $R_{t}[j, y]$; store the value of $M[i, j]$ into $M_{r}[i, j]$ for each value $i$ in $I_{y}$ as well;}
\item{set $g$ to be $y-1$;}
\item{compute the $(j, g, yw+1)$ partial right difference vector $U_g$ and the $(yw+1, j-1)$ partial deviation $d_g$ for $((g+1)w, j)$, store the integer encoded by $U_g$ and $d_g$ into $R_{p}[j, g]$;}
\item{decrement $g$ by $1$;}
\item{if $g \geq 0$ holds, go to step 10;}
\item{decrement $y$ by $1$;}
\item{if $t \geq 0$ holds, go to step 8;}
\item{if $j$ is the last position of $I_s$, update $L_t$ to store the integer encoded by $(i, s)$ left difference vector into $L_t[i, s]$ for each $i$ that satisfies $1 \leq i \leq sw$;}
\item{if $j$ is the last position of $I_s$, for each $q$ that satisfies $0 \leq q < s$, precompute the content of the locations in $C_p$ for pair $(q, s)$;}
\item{increment $j$ by $1$;}
\item{if $j \leq n$ holds, go to step 3;}
\item{use the trace-back algorithm developed in Section \ref{sec0} to find the base pairs in the secondary structure and output them.}
\end{enumerate}

In steps 4, 5 and 6, the algorithm exhaustively enumerates all integers $q$ that satisfy $sw+1 \leq q < j$ for each $i$ and computes the maximum value of $M[i, q]+M[q+1, j]$ and stores the
value into $M_{r}[i, j]$ if $i$ is in $I_s$. The value is stored in an array $N$ if $i$ is not in $I_s$. For each $h$ that satisfies $1 \leq h \leq s-1$, the $(j, h, sw+1)$ partial right difference vector $U_h$ and $(sw+1, j-1)$ partial deviation $d_h$ for $(hw, j)$ can both be computed from $N$. The integer encoded by $U_h$ and $d_h$ are then stored into table $R_p$ at $R_{p}[j, h]$.

In step 8, the algorithm obtains the $(j, y, (y+1)w+1)$ partial right difference vector and the value of the corresponding $((y+1)w+1, j-1)$ partial deviation for $((y+1)w, j)$ by querying the table $R_{p}$ with $j$ and $y$, it then computes the $(j, t)$ right difference vector using the algorithm in the proof of Lemma \ref{lm7}. This step finalizes the computation of $M[i, j]$ for each $i$ that is in $I_{y}$.

In step 10, the algorithm queries $R_p$ with $j$ and $g$ to obtain the integer encoded by the $(j, g, (y+1)w+1)$ partial right difference vector $U_g$ and the $((y+1)w+1, j-1)$ partial deviation $d_{y}$ for $((g+1)w, j)$. It then queries $R_t$ with $j$ and $y$ to obtain the integer encoded by $(j, y)$ right difference vector and $L_t$ with $(g+1)w$ and $y$ to obtain the integer encoded by the $((g+1)w, j)$ left difference vector. We use $e_y$ and $f_g$ to denote these two integers respectively. It then queries $C_p$ with $g$, $y$, $e_y$, and $f_g$ to obtain the integer encoded by the $(j, g, y)$ partial central difference vector $V_g$ and the corresponding $(yw+1, (y+1)w)$ partial deviation $d_{a}$ for $((g+1)w, j)$. We use $u_g$, $v_g$ to denote the integers encoded by $U_g$ and $V_g$ respectively, the algorithm computes two other integers as follows.
\begin{equation}
d_1=d_{y}
\end{equation}
\begin{equation}
d_2=M_r[(g+1)w, (y+1)w]+M_r[(y+1)w+1, j]-M[(g+1)w, j-1]+d_{a}+n
\end{equation}
It then queries $U_p$ to obtain the two integers stored at $U_{p}[u_{g}, d_{1}, v_{g}, d_{2}]$. One of these two integers is the integer encoded by the $(j, g, yw+1)$ partial right difference vector and the other one is the corresponding $(yw+1, j-1)$ partial deviation for $((g+1)w, j)$. We store both of them into
$R_p$ at $R_p[j, g]$.

In steps 13 and 14, the algorithm updates the table $L_t$ if $j$ is in the last position of $I_s$. For each $i$ that satisfies $1 \leq i \leq sw$, the integer encoded by the $(i, s)$ left difference vector is stored in $L_t$ at $L_t[i, s]$. In addition,  since the values of all $M[i, j]$'s for each $j$ in $I_s$ have been computed, the algorithm performs the precomputation needed to determine the content of the locations associated with each pair $(q, s)$ in $C_p$, where $0 \leq q \leq s$.

\begin{theorem}
\rm
\label{th4}
Given an RNA sequence of length $n$, our algorithm solves the basic RNA folding problem in $O(\frac{n^3}{\log_{2}^{2}{n}})$ time and $O(n^{2.5})$ space.

\begin{proof}
We show that our algorithm correctly solves the basic RNA folding problem in $O(\frac{n^3}{\log_{2}^{2}{n}})$ time and $O(n^{2.5})$ space. Steps 4, 5 and 6 in the algorithm guarantee the correctness of the computed values for $M[i, j]$'s and $P[i, j, sw+1, j-1]$'s since an exhaustive search is performed on all integers between $sw+1$ and $j-1$ in these steps.

The correctness of the $(j, y)$ right difference vector computed in step 8 comes from Lemma \ref{lm7} and the fact that the integer encoded by the $(j, y, (y+1)w+1)$ partial right difference vector and the value of the corresponding $((y+1)w+1, j-1)$ partial deviation for $((y+1)w, j)$ have been correctly computed when $I_{y+1}$ is processed.

Lemma \ref{lm5} and Lemma \ref{lm6} guarantee that the integer encoded by the $(j, g, yw+1)$ partial right difference vector and the corresponding $(yw+1, j-1)$ partial deviation for $((g+1)w, j)$ can be correctly computed based on a few intermediate computational results. These results include the integer encoded by the $(j, g, (y+1)w+1)$ partial right difference vector, the $((y+1)w+1, j-1)$ partial deviation for $((g+1)w, j)$, the integer encoded by the $((g+1)w, y)$ left difference vector, the integer encoded by the $(j, y)$ right difference vector, and
the values of $M[(s+1)w, (y+1)w]$, $M[(y+1)w+1, j]$, and $M[(g+1)w, j-1]$. It is straightforward to see that all these results have been correctly computed and stored in memory before step 10 is executed. Finally, steps 13 and 14 correctly update $L_t$ and $C_p$ when $j$ is in the last position of $I_s$. The correctness of the algorithm thus follows.

From Lemma \ref{lm6}, the precomputation of $U_p$ needs $O(n^{2}w2^{2w})$ time. From Lemma \ref{lm5}, the precomputation of $C_p$ for a given pair of aggregate intervals need $O(w^{3}2^{2w})$ time. Since there are $O(\frac{n^{2}}{w^{2}})$ pairs for which $C_p$ needs to be computed, the total amount of time needed to precompute $C_p$ is thus at most $O(n^{2}w2^{2w})$. For each $j$, the algorithm needs $O(nw)$ time in steps 4, 5 and 6.  Step 8 is executed only once for each aggregate interval for each $j$ and it needs time $O(w^2)$ from Lemma \ref{lm7}. Step 10 can be executed in $O(1)$ time and it is executed for $O(\frac{n^{2}}{w^{2}})$ times for each $j$. Step 13 can be completed in time $O(n)$ for each $j$. From Theorem \ref{th2}, the last step needs $O(n^2)$ time. The total amount of computation time $T(n)$ needed by the algorithm thus must satisfy
\begin{equation}
T(n) \leq A(n^{2}w^{2}2^{2w}+n^{2}w+\frac{n^{3}}{w^2}+n^{2})
\end{equation}
where $A$ is some positive constant. Let $w=\frac{1}{4}\log_{2}{n}$ and the time complexity of the algorithm is thus $O(\frac{n^{3}}{\log_{2}^{2}{n}})$.

For space complexity, $L_t$, $R_t$ and $R_p$ need $O(\frac{n^2}{w})$ space; $U_p$ needs $O(n^{2}2^{2w})$ space and $C_p$ needs $O(\frac{n^{2}2^{2w}}{w^{2}})$ space; $M_r$ needs $O(n^2)$ space. It is straightforward to see that the space complexity of the algorithm is $O(n^{2.5})$ when $w=\frac{1}{4}\log_{2}{n}$. The theorem thus follows.
\end{proof}
\end{theorem}

It is worth pointing out that we can reduce the space complexity of the algorithm to $O(n^{2+\epsilon})$ for any $\epsilon > 0$ by setting $w=\frac{1}{2}\epsilon \log_{2}{n}$. Although this does not change the asymptotic time complexity of the algorithm, a large hidden constant factor is introduced to the time complexity and doing this thus may not be desirable in practice.

\section{Extend to the Context-Free Language Recognition Problem}

In this section, we consider the worst-case space complexity of the context-free language recognition problem. The goal of the problem is to determine whether a string is in the language defined by a context-free grammar. The problem can be solved in time $O(gn^{3})$ with the well known CYK algorithm, where $g$ is the number of nonterminals in the grammar and $n$ is the length of the string. This algorithm shares the same recursive structure as the traditional dynamic programming algorithm for the basic RNA problem.

In \cite{valiant}, a subcubic algorithm is developed to solve the problem based on a fast algorithm for boolean matrix multiplication. Since the algorithm is complex and difficult to implement in practice, a few other algorithms that can solve the problem in $o(n^3)$ time were developed afterwards. For example, in \cite{graham}, an $O(\frac{n^3}{\log_{2}{n}})$ time algorithm is developed to solve the problem using the Four-Russians Technique. In \cite{wojceich}, Compression techniques are used to obtain an $O(\frac{n^3}{\log_{2}^{2}{n}})$ time algorithm for the problem. In \cite{lee}, a reduction from the boolean matrix multiplication problem to the problem is established and it is shown that the time complexities of the two problems are in fact inherently related.

Although the worst-case time complexity of the problem has been improved to $o(n^3)$ or even subcubic time. The worst-case space complexity for efficiently solving the problem remains to be $O(n^2)$. In \cite{alt}, a lower bound of $O(\log_{2}{n})$ is established for the space complexity of this problem. However, it remains unknown whether the problem can be efficiently solved in $o(n^2)$ space. As has been observed in previous work, the dynamic programming tables for this problem contain only $0$ or $1$'s. The memory efficient algorithm we have developed in Section \ref{sec1} can thus be slightly modified to efficiently solve the problem in $O(\frac{gn^{2}}{\log_{2}{n}})$ space.

\begin{corollary}
\rm
The context-free language recognition problem can be solved in time $O(\frac{gn^{3}}{\log_{2}{n}})$ while using only $O(\frac{gn^{2}}{\log_{2}{n}})$ space, where $g$ is the number of nonterminals in the grammar and $n$ is the length of the string.
\end{corollary}

\section{Conclusions}
In this paper, we develop new algorithms for the basic RNA folding problem. Based on the Four-Russians Technique, we show that the problem can be solved with a traditional dynamic programming algorithm in time $O(\frac{n^{3}}{\log_{2}{n}})$ while using only $O(\frac{n^{2}}{\log_{2}{n}})$ space, where $n$ is the length of the sequence. In addition, we show that the Four-Russians Technique can be further utilized to solve the problem in $O(\frac{n^{3}}{\log_{2}^{2}{n}})$ time using only traditional dynamic programming techniques. Our work also improves the worst-case space complexity for efficienlty solving the context-free language recognition problem to $O(g\frac{n^2}{\log_{2}{n}})$, where $g$ is the number of nonterminals in the grammar and $n$ is the length of the string.

In \cite{valiant}, the Valiant's technique is used to solve the context-free language recognition problem in subcubic time. A question therefore arises naturally on whether subcubic time solutions also exist for the basic RNA folding problem. The techniques developed in \cite{valiant} are based on boolean matrix multiplications and thus cannot be directly applied to the basic RNA folding problem. Although a reduction from the basic RNA folding problem to the stochastic context-free parsing is available \cite{durbin}, an efficient reduction that reduces the problem to the context-free language recognition problem is still unavailable. The worst-case time complexity of this problem thus remains to be an interesting problem and deserves additional investigations in the future.

\end{document}